  \documentstyle[11pt,epsfig]{article}
  
\newcommand{\msbar}{\overline{\mbox{\scriptsize MS}}}
\newcommand{\MSbar}{\overline{\mbox{MS}}}
\newcommand{\RI}{\mbox{RI}}
\newcommand{\ri}{\mbox{\scriptsize RI}}

\newcommand{\Tr}{\mbox{Tr}\;}

\newcommand{\bc}{\begin{center}}
\newcommand{\ec}{\end{center}}
\newcommand{\be}{\begin{equation}}
\newcommand{\ee}{\end{equation}}
\newcommand{\bea}{\begin{eqnarray}}
\newcommand{\eea}{\end{eqnarray}}
\newcommand{\ba}{\begin{eqnarray}}
\newcommand{\ea}{\end{eqnarray}}
\newcommand{\brr}{\begin{array}}
\newcommand{\err}{\end{array}}

\def\dfrac#1#2{{\displaystyle {#1 \over #2}}}

\newcommand{\simge}{\ \lower-
1.2pt\vbox{\hbox{\rlap{$>$}\lower5pt
\vbox{\hbox{$\sim$}}}}\ }

\def\spose#1{\hbox to 0pt{#1\hss}}
\def\ltapprox{\mathrel{\spose{\lower 3pt\hbox{$\mathchar"218$}}
 \raise 2.0pt\hbox{$\mathchar"13C$}}}
\def\gtapprox{\mathrel{\spose{\lower 3pt\hbox{$\mathchar"218$}}
 \raise 2.0pt\hbox{$\mathchar"13E$}}}
\def\inapprox{\mathrel{\spose{\lower 3pt\hbox{$\mathchar"218$}}
 \raise 2.0pt\hbox{$\mathchar"232$}}}

\begin{document}
\pagestyle{empty}
\vspace{-0.6in}
\begin{flushright}
BUHEP-01-14\\
\end{flushright}
\vskip 1.5in

\centerline{\large {\bf{Light Quark Masses with Overlap Fermions
in Quenched QCD}}}
\vskip 0.6cm
\centerline{\bf L.~Giusti, C.~Hoelbling, C.~Rebbi}
\vskip 0.2cm
\centerline{ Department of Physics - Boston University}
\centerline{590 Commonwealth Avenue, Boston MA 02215, USA}
\vskip 2.5cm
\begin{abstract}
We present the results of a computation of the sum of the strange and
average up-down quark masses with overlap fermions in the quenched
approximation.  Since the overlap regularization preserves chiral 
symmetry at finite cutoff and volume, no additive quark mass
renormalization is required and the results are ${\cal O}(a)$
improved. Our simulations are performed at $\beta=6.0$ and volume
$V = 16^3\times 32$, which correspond to a lattice cutoff of $\sim
2$~GeV and to an extension of $\sim 1.4$~fm. The 
logarithmically divergent renormalization constant has been computed 
non-perturbatively in the RI/MOM scheme.  By using the $K$-meson mass
as experimental input, we obtain 
$(m_s + \hat m)^{\ri}(2\; \mbox{GeV}) = 120(7)(21)$~MeV, which corresponds 
to $m_s^{\msbar}(2\; \mbox{GeV})
= 102(6)(18)$~MeV if 
continuum perturbation theory and $\chi$PT are used. By using the GMOR relation 
we also obtain $\langle \bar \psi \psi \rangle^{\msbar}(2\, \mbox{GeV})/N_f
= - 0.0190(11)(33)\, \mbox{GeV}^{3} = - [267(5)(15)\, \mbox{MeV}]^3$.
\end{abstract}
\vfill
\pagestyle{empty}\clearpage
\setcounter{page}{1}
\pagestyle{plain}
\newpage
\pagestyle{plain} \setcounter{page}{1}

\newpage

\section{Introduction}
Quark masses are fundamental parameters of the Standard Model which
cannot be measured directly by experiment, since quarks are confined
into hadrons. If defined as effective couplings in the Lagrangian,
their values can be determined by comparing a theoretical
calculation of a given physical quantity (sensitive to quark
masses) with the corresponding experimental value.  As a consequence
quark masses depend on the renormalization scheme and scale, as well 
as on the fundamental action.

At present the most precise values of light-quark mass ratios (which
are scheme and scale independent for mass independent renormalization
schemes) are extracted by comparing $K$- and $\pi$-meson mass ratios
with predictions from chiral perturbation theory ($\chi$PT)~\cite{Gasser:1982ap}.
A detailed analysis gives~\cite{Leutwyler:1996qg}
\be\label{eq:chipt}
\frac{m_u}{m_d} = 0.553 \pm 0.043\, , \qquad \frac{m_s}{\hat m} = 24.4 \pm 1.5
\ee
where $\hat m = (m_u+m_d)/2$. The absolute scale cannot be fixed by
$\chi$PT.  It can be determined by comparing non-perturbative lattice
QCD computations \cite{Allton:1994ae}-\cite{AliKhan:2000mw} or
phenomenological estimates \cite{narison,pich} with experimental
results.

In the last few years much effort has been devoted within the lattice
community to obtain a precise determination of the light quark masses in
the quenched approximation. Major improvements came with the use of
non-perturbative (NP) renormalization techniques for renormalizing the
bare quark masses and with the implementation of ${\cal O}(a)$
improved actions and operators
\cite{Gimenez:1999uv,Becirevic:1998yg,Aoki:1999mr,Garden:2000fg} (for
a recent review see \cite{vittorio_rev}). First results from
unquenched simulations have also been reported
\cite{Eicker:1999sy,AliKhan:2000mw,vittorio_rev}.

In this paper we present the results of the first fully
non-perturbative computation of $(m_s+\hat m)$ with overlap fermions.
In this fermionic regularization, flavor and chiral
symmetries are preserved at finite lattice spacing and finite
volume. As a consequence no additive quark mass renormalization is
required and no parameters have to be fine tuned in order to compute
${\cal O}(a)$ improved masses and matrix elements. To avoid
uncertainties due to lattice perturbation
theory, we compute the logarithmic divergent renormalization constant
non-perturbatively in the RI/MOM scheme \cite{NP}.  By comparing the
experimental $K$-meson mass with the value obtained in our simulations
and after a careful analysis of the systematic uncertainties we find
as our main result
\be\label{eq:mainri}
(m_s + \hat m)^{\ri}(2\; \mbox{GeV}) = 120 \pm 7 \pm 21\, \mbox{MeV}
\ee
This value corresponds to
\be\label{eq:msbar}
m_s^{\msbar}(2\; \mbox{GeV}) = 102 \pm 6 \pm 18\, \mbox{MeV}
\ee
if next-to-next-to-leading order (N$^2$LO) continuum perturbative
results and Eq.~(\ref{eq:chipt}) are used.

We also report results for the chiral condensate
$\langle \bar \psi \psi \rangle$ which we compute 
from the Gell-Mann, Oakes and Renner (GMOR) relation.
Our best determination is
\be\label{eq:mainri_cond}
\frac{1}{N_f}\langle \bar \psi \psi \rangle^{\ri}(2\, \mbox{GeV}) = -\, 0.0167 \pm
0.0010 \pm 0.0029\, \mbox{GeV}^3
\ee
which corresponds to
\be\label{eq:msbar_cond}
\frac{1}{N_f}\langle \bar \psi \psi \rangle^{\msbar}(2\, 
\mbox{GeV}) = -\, 0.0190 \pm
0.0011 \pm 0.0033\, \mbox{GeV}^3 = -\, [267 \pm 5 \pm 15\, \mbox{MeV}]^3
\ee
We have also computed the chiral condensate directly. Even if it requires 
a severe chiral extrapolation, the direct determination gives results
consistent with Eqs.~(\ref{eq:mainri_cond}), (\ref{eq:msbar_cond}).

The first error in Eqs.~(\ref{eq:mainri})-(\ref{eq:msbar_cond}) 
represents the statistical
error, obtained with the jackknife method.
The second error represents our estimate of systematic effects.
It should be noted that, while the systematic error due to
quenching alone (defined as the quenching error in the limit of
zero lattice spacing and infinite volume) should be the same in all
fermionic regularizations, discretization errors
do depend on the fermionic lattice
action and are likely to be smaller in the overlap formulation,
because of its good chirality properties.  The very good agreement
of our results with the current lattice world averages~\cite{vittorio_rev}
provides further confirmation that the overlap formulation
is a suitable regularization for large-scale phenomenological computations.

This paper is organized as follows: in Sec.~\ref{sec:Action} we set
our notation and define the renormalized quark masses and chiral
condensate in the overlap regularization; in Sec.~\ref{sec:numerology}
we give details about the simulation and some of the results for the
meson masses and matrix elements; in Sec.~\ref{sec:Results} we present
our main results and discuss their systematic errors;
Sec.~\ref{sec:conclusions} is devoted to our concluding remarks.

\section{Quark Masses and Chiral Condensate with the
Overlap Action}\label{sec:Action}

The QCD lattice action in the overlap regularization reads \cite{neub1}
\be
S = \frac{6}{g_0^2}\sum_{P} \bigg[ 1 - \frac{1}{6}
\Tr \Big[ U_P + U_P^{\dagger} \Big] \bigg]  +
 \bar{\psi} \left[\Big(1-\frac{1}{2\rho}a M \Big)D + M\right] \, \psi
\label{eq:sg_QCD}
\ee
where, in standard notation, $U_P$ is the Wilson plaquette,
$g_0=\sqrt{6/\beta}$ is the bare coupling constant,
$\psi$ and $\bar\psi$ carry implicit color, spin and flavor indices,
and $M$ is a diagonal matrix of bare masses $(m_1,m_2,\dots)$
in flavor space.
$D$ is the Neuberger-Dirac operator defined as
\ba\label{eq:opneub}
D &=&  \frac{\rho}{a}\left( 1 + V \right) =
\frac{\rho}{a}\left( 1 + X\frac{1}{\sqrt{X^\dagger X}}\right)\nonumber\\
X &=& D_W -\frac{1}{a}\rho
\ea
where
\be\label{eq:WDO}
D_W = \frac{1}{2} \gamma_\mu (\nabla_\mu + \nabla^*_\mu)
-\frac{r}{2} a \nabla^*_\mu\nabla_\mu
\ee
is the Wilson-Dirac operator, $0<r\leq 1$ and
$0<\rho< 2r$.  In our calculations we used $r=1$.  $\nabla_\mu$ and
$\nabla^*_\mu$ in Eq.~(\ref{eq:WDO}) are the forward and backward lattice
covariant derivatives, defined by
\ba
\nabla_\mu \psi(x) & = & \frac{1}{a}\Big[U_\mu(x) \psi(x + a \hat \mu) -
\psi(x)\Big]\label{eq:derivative1}\\
\nabla_\mu^* \psi(x) & = &
\frac{1}{a}\Big[ \psi(x) - U^{\dagger}_\mu(x-a \hat \mu)
\psi(x-a\hat\mu) \Big] \label{eq:derivative2}
\ea
where $U_\mu(x)$ are the lattice gauge links. The fermionic
operator of the overlap action satisfies the Ginsparg-Wilson relation \cite{GW}
\be\label{eq:GW}
\gamma_5 D + D \gamma_5 = \frac{a}{\rho} D \gamma_5 D
\ee
which implies an exact continuous symmetry of the action in the massless
limit \cite{luscher}. This symmetry may be interpreted as a lattice form
of chiral invariance at finite cutoff
\be\label{eq:luscher_new}
\delta \psi = \hat \gamma_5 \psi\; , \qquad
\delta \bar \psi = \bar \psi \gamma_5
\ee
where $\hat \gamma_5$ is defined as
\be
\hat \gamma_5 = \gamma_5\Big(1-\frac{a}{\rho}D\Big)
\ee
and satisfies
\be
\hat \gamma_5^\dagger = \hat\gamma_5\; , \qquad \hat \gamma_5^2 = 1
\ee
The anomaly is recovered from the variation of the measure under the
rotations in Eq.~(\ref{eq:luscher_new}) \cite{luscher,hln}.

The invariance of the action under non-singlet chiral
transformations, defined including a flavor group generator
in Eq.~(\ref{eq:luscher_new}), forbids mixing among operators of
different chirality \cite{hasenfratz2} and therefore:
\begin{itemize}
\item No additive quark mass renormalization is required. The 
      quark mass which enters the vector and
      axial Ward identities is the bare parameter $m(a)$
      in the action of Eq.~(\ref{eq:sg_QCD}).
\item Masses and matrix elements are affected only by
      ${\cal O}(a^2)$ discretization errors.  No fine tuned 
      parameters are  required to remove ${\cal O}(a)$ effects.
\item The chiral condensate does not require subtractions of 
      power divergent terms (in the chiral limit).
\end{itemize}
The renormalized quark mass is defined as
\be
\bar m(\mu) = \lim_{a\rightarrow 0} Z_m(a \mu) m(a)
\ee
where $Z_m(a \mu)$ is a logarithmically divergent
renormalization constant which has to be fixed, for a given
scale $\mu$, in a given renormalization scheme.
It is worth noting that even if $m(a)$ is the
bare parameter which enters the fundamental action in Eq.~(\ref{eq:sg_QCD}),
its relation to a given experimental result is fixed by
a non-perturbative lattice QCD calculation and therefore
its value is determined up to ${\cal O} (a^2)$ terms only.

The bare chiral condensate is defined as
\be\label{eq:cond_def}
\chi(a) \equiv \lim_{m \rightarrow 0}\frac{1}{N_f}
\langle \bar \psi(0) [(1-\frac{a}{2\rho} D) \psi](0) \rangle
\ee
where $m$ is a common mass given to the light quarks.
It satisfies the integrated non-singlet chiral Ward identity
\be\label{eq:bella}
\frac{1}{N_f}
\langle \bar \psi(0) [(1-\frac{a}{2\rho} D) \psi](0) \rangle
= \, m \sum_{x} \langle P(x) P^c(0)\rangle
\ee
where
\ba
P(x) & = & \bar{\psi}_1(x) \gamma_5 [(1-\frac{a}{2\rho} D) 
\psi_2](x)\label{eq:pseudoscalar1}\\
P^c(x) & = & \bar{\psi}_2(x) \gamma_5 [(1-\frac{a}{2\rho} D) 
\psi_1](x)\label{eq:pseudoscalar2}
\ea
correspond the non-singlet pseudoscalar density with degenerate quarks
($m_1=m_2$) and its conjugate. For non-zero mass
the chiral condensate is still divergent and it behaves as
\be\label{eq:cond_div}
\frac{1}{N_f}\langle \bar \psi(0) [(1-\frac{a}{2\rho} D) \psi](0) \rangle =
\chi(a) + \beta_\chi \frac{m(a)}{a^2}\; ,
\ee
where we have taken into account that chiral symmetry forces
the coefficient of the linear divergence to be zero.

By writing the correlation function
$\langle P(x) P^c(0) \rangle$ as a time-ordered product and by inserting a
complete set of states in standard fashion we can also write
\be
\label{eq:wip}
\chi(a) =
-\, \lim_{m \rightarrow 0}
\dfrac{m}{M_P^2}
\Big \vert \langle 0 \vert P \vert P \rangle \Big \vert^2\; ,
\ee
where $M_P$ is the mass of the pseudoscalar state $\vert P \rangle$.
If we use
\be\label{eq:fp}
2 m |\langle 0 | P | \pi\rangle| =  f_{P} M^2_{P}
\ee
where $f_{P}$ is the corresponding pseudoscalar decay constant, we arrive
to the familiar GMOR relation
\be\label{eq:GMOR_true}
\chi(a) =
- \lim_{m \rightarrow 0} \frac{f^2_{P} M^2_{P}}{4 m}
\ee

The renormalized chiral condensate is defined as
\be
 \frac{1}{N_f}\langle \bar \psi \psi\rangle(\mu) = \lim_{a\rightarrow 0} Z_S(a \mu)
\chi(a)
\ee
where $Z_S(\mu a)$ can be chosen to be the renormalization constant of the
corresponding non-singlet scalar density which, thanks to the flavor
symmetry, satisfies $Z_S(\mu a) = 1/Z_m(\mu a)$.
In principle $Z_S(\mu a)$ can
be computed in perturbation theory \cite{vicari,capgiu}, but
uncertainties due to higher order terms can be avoided
using non-perturbative renormalization procedures 
\cite{NP,luscher_np,lambdaqcd}. 
The implementation of the RI/MOM technique \cite{NP} is straightforward
in the overlap regularization, and it allows one to compute ${\cal O} (a)$ 
improved renormalization constants for generic composite operators.
In the following we will use the numerical value of 
$Z^{\ri}_S(\mu a)$ we have obtained in Ref.~\cite{noi_np}
(for an alternative approach see Ref.~\cite{HJL_NP2}). 

\section{Numerical Details}\label{sec:numerology}
We performed our simulation in quenched QCD with $\beta=6.0$ and
$V=16^3 \times 32$. We used a sample of
$54$ gauge configurations, generated with the standard Wilson gluonic
action of Eq.~(\ref{eq:sg_QCD}), which we retrieved from the
repository at the ``Gauge Connection'' \cite{GaugeConn}. We computed
overlap propagators from a local source for bare quark masses
$ma=0.040,0.055,0.070,0.085,0.100$ and $\rho=1.4$
(and, as already mentioned, $r=1.0$). For the calculation
of the propagators, we used an optimal rational approximation
to the sign of the Hermitian Wilson operator, as proposed in
Ref.~\cite{Neuberger:1998my,bob_cond}, after explicit evaluation of the
contributions from the lowest eigenvectors of $X^\dagger X$, and nested 
multi-conjugate
gradient inversions.  Details of the numerical implementation
will be presented in Ref.~\cite{noi_np}.
 From the propagators, we computed in the standard manner the
two-points correlation functions\footnote{Analogous correlation functions
have been computed by the authors of Ref.~\cite{Liu}. A direct comparison 
with our results is not possible because we used different simulations 
parameters.}
\ba\label{eq:scal_vec}
G_{SS}(t) & = & \sum_{x}\langle S(x,t)S^{c}(0,0) \rangle\\
G_{PP}(t) & = & \sum_{x}\langle P(x,t)P^{c}(0,0) \rangle\\
G_{\nabla AP}(t) & = & \sum_{x}\langle \bar \nabla_0 A_0(x,t)P^{c}(0,0)\rangle
\ea
where $P(x,t)$ and $P^c(x,t)$ have been defined in 
Eqs.~(\ref{eq:pseudoscalar1}) and (\ref{eq:pseudoscalar2}),
\ba
S(x)  & = & \bar{\psi}_1(x) [(1-\frac{a}{2\rho} D)\psi_2](x)\\
A_\mu(x)  & = & \bar{\psi}_1(x) \gamma_\mu \gamma_5[(1-\frac{a}{2\rho} D)
\psi_2](x)\;
\ea
$S^c(x)$ is defined analogously to Eq.~(\ref{eq:pseudoscalar2})
and $\bar \nabla_0$ denotes the symmetric lattice derivative in the time
direction. To improve statistics $G_{SS}(t)$ and $G_{PP}(t)$
have been symmetrized around $t=T/2$  ($T=N_t=32$).
We estimated the errors by a jackknife procedure, blocking the data in
groups of three configurations, and we checked that blocking in groups
of different size did not produce relevant changes in the error estimates.
\begin{figure}[htb]
\begin{center}
\begin{tabular}{cc}
\mbox{\epsfig{file=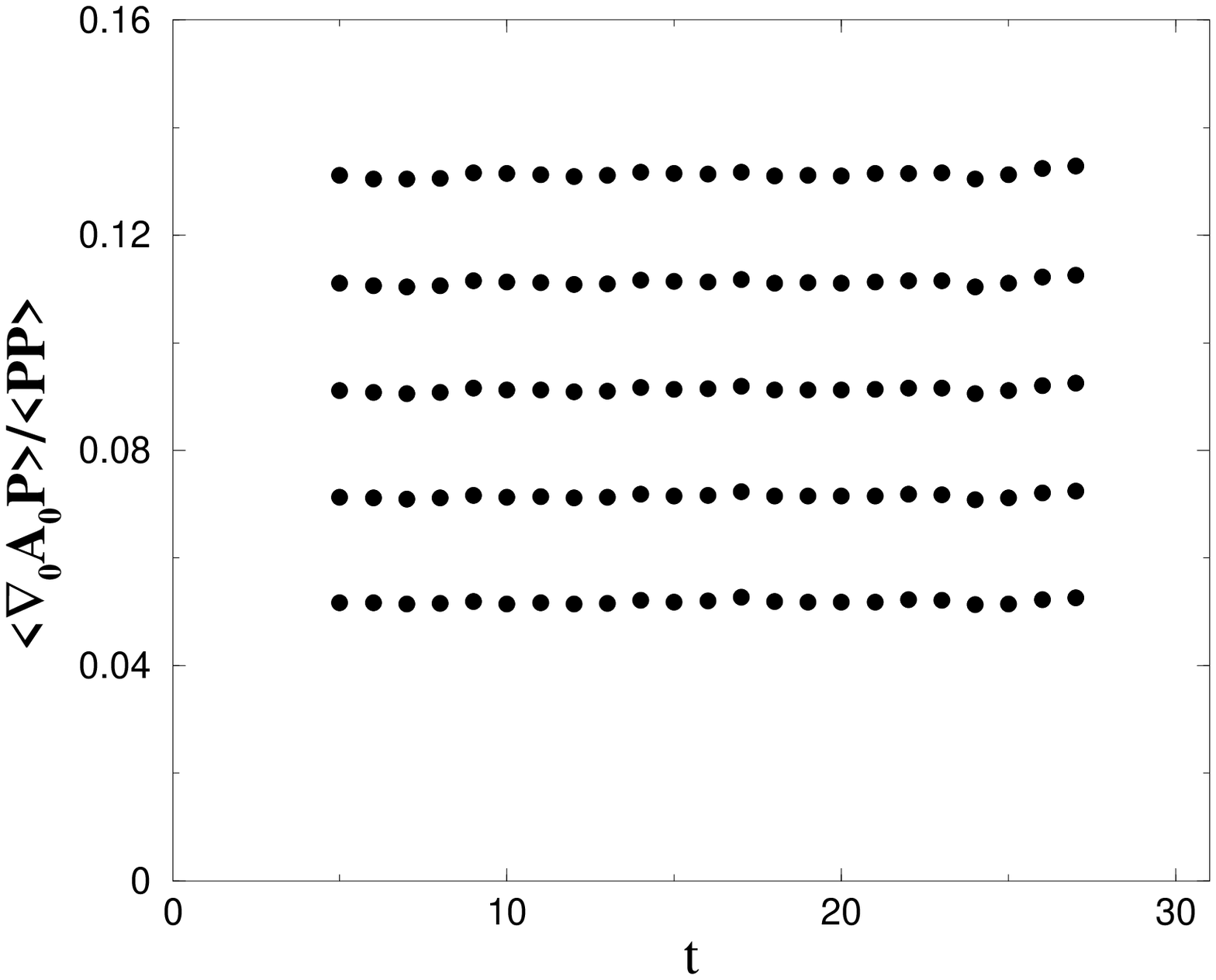,height=6.0cm,width=8.0cm,angle=0}} &
\mbox{\epsfig{file=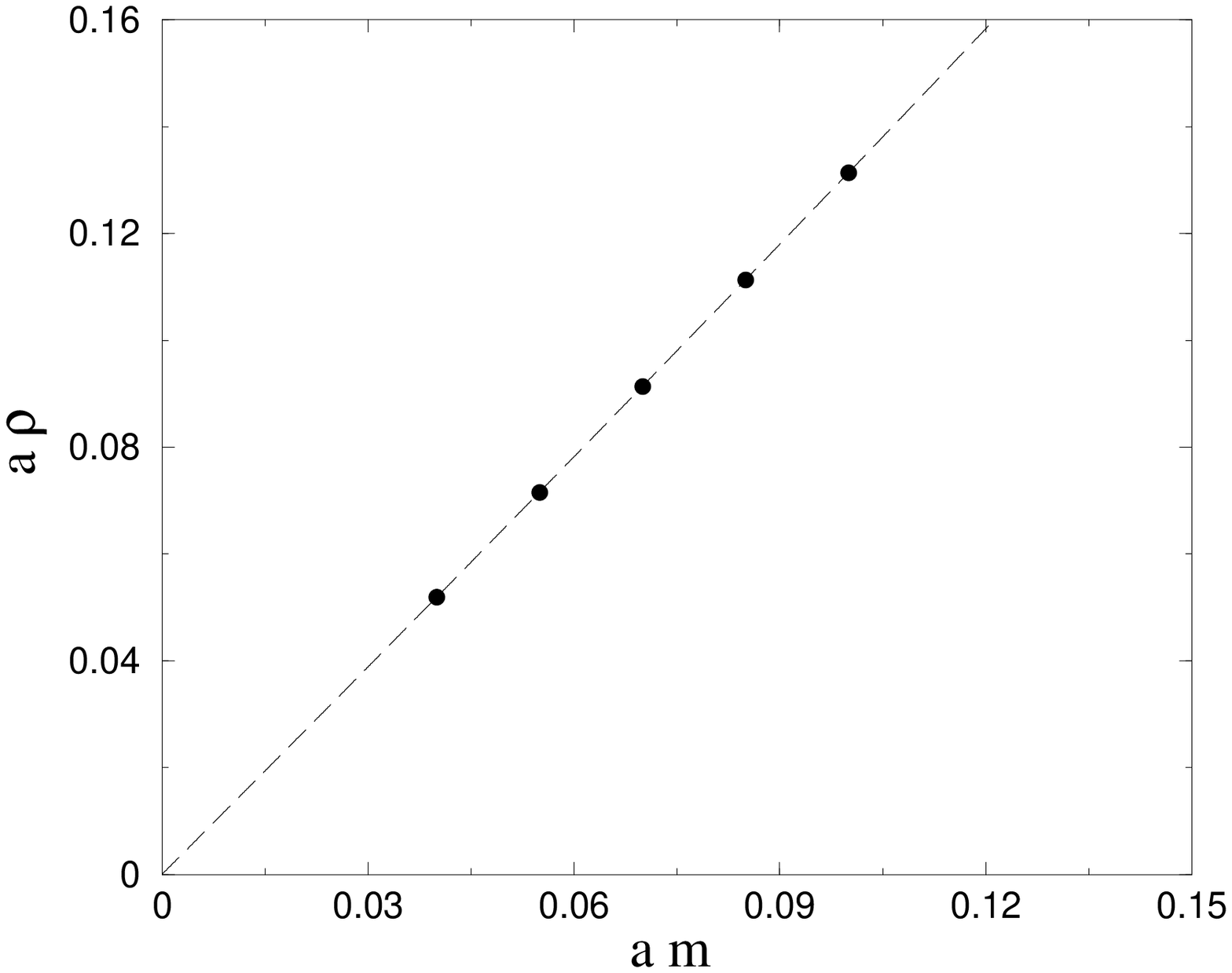,height=6.0cm,width=8.0cm,angle=0}} \\
\end{tabular}
\caption{Left: $G_{\nabla AP}/G_{PP}$ vs.~$t$ for all masses
which have been simulated. Right: $(a \rho)$, obtained by fitting the data
in the plot at left, as a function of the bare quark mass.  The dashed line
represents the result of a quadratic fit (see text).
\label{fig:axials}}
\end{center}
\end{figure}
In the first plot of Fig.~\ref{fig:axials} we show the ratio
\be
\rho(t) \equiv \frac{G_{\nabla AP}(t)}{G_{PP}(t)}
\ee
as a function of $t$ for all simulated masses.
Once $\rho$ has been fitted to a constant
in the time interval $t_1-t_2=5-27$, a quadratic fit of the
results
\be
(a \rho) = {\cal A}_\rho + {\cal B}_\rho (a m) + {\cal C}_\rho (a m)^2
\ee
gives\footnote{From the coefficient ${\cal B}_\rho$ one can derive the value
of the renormalization constant $Z_A$ of the ``local'' axial current. A 
detailed analysis will be presented
in Ref.~\cite{noi_np}.} (see the second plot in Fig.~\ref{fig:axials})
\be
{\cal A}_\rho = -0.00002(7) \qquad {\cal B}_\rho = 1.286(3) \qquad {\cal C}_\rho = 0.277(12)
\ee
where the quoted errors are statistical only.
\begin{figure}[htb]
\begin{center}
\begin{tabular}{cc}
\mbox{\epsfig{file=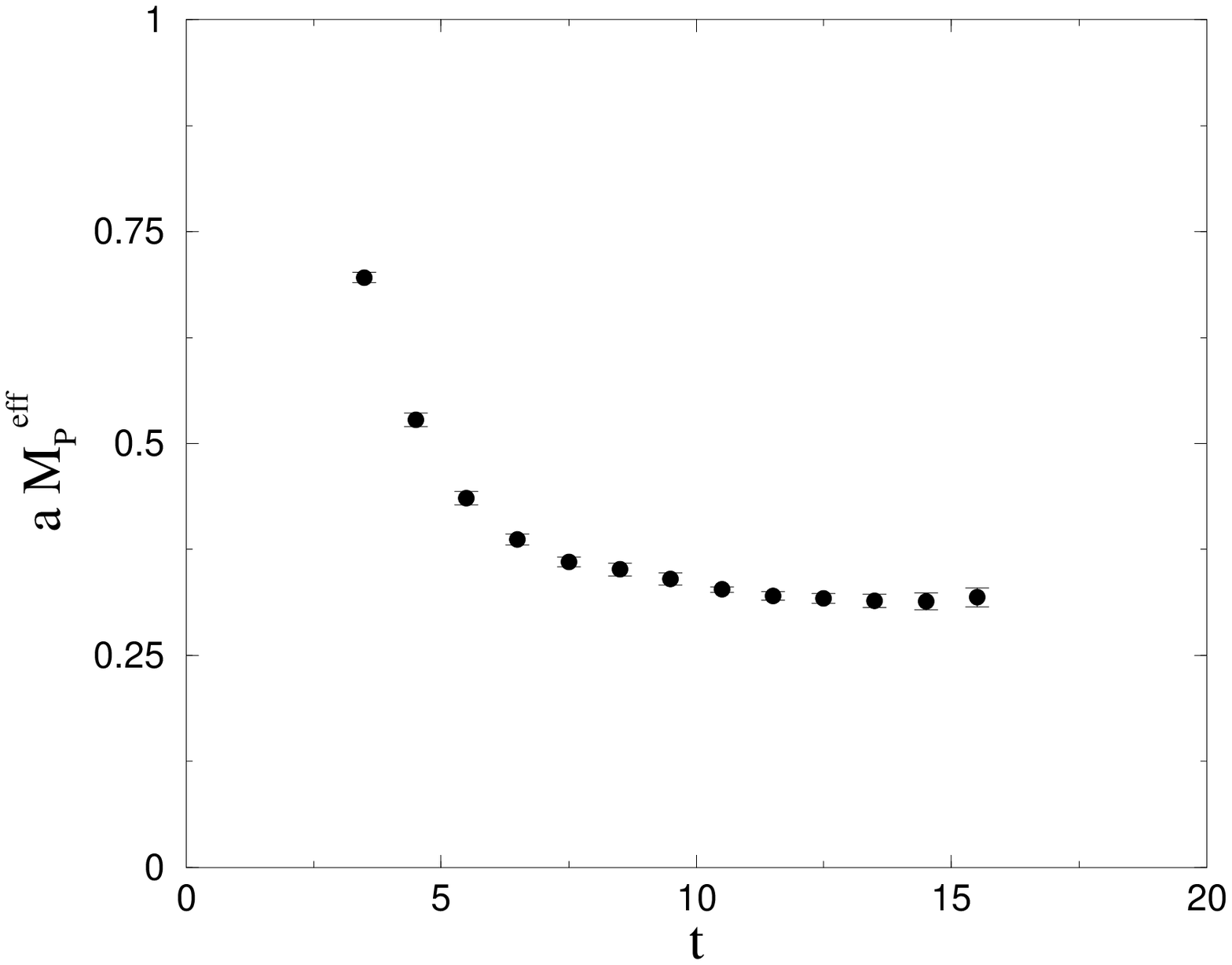,height=6.0cm,width=8.0cm,angle=0}} &
\mbox{\epsfig{file=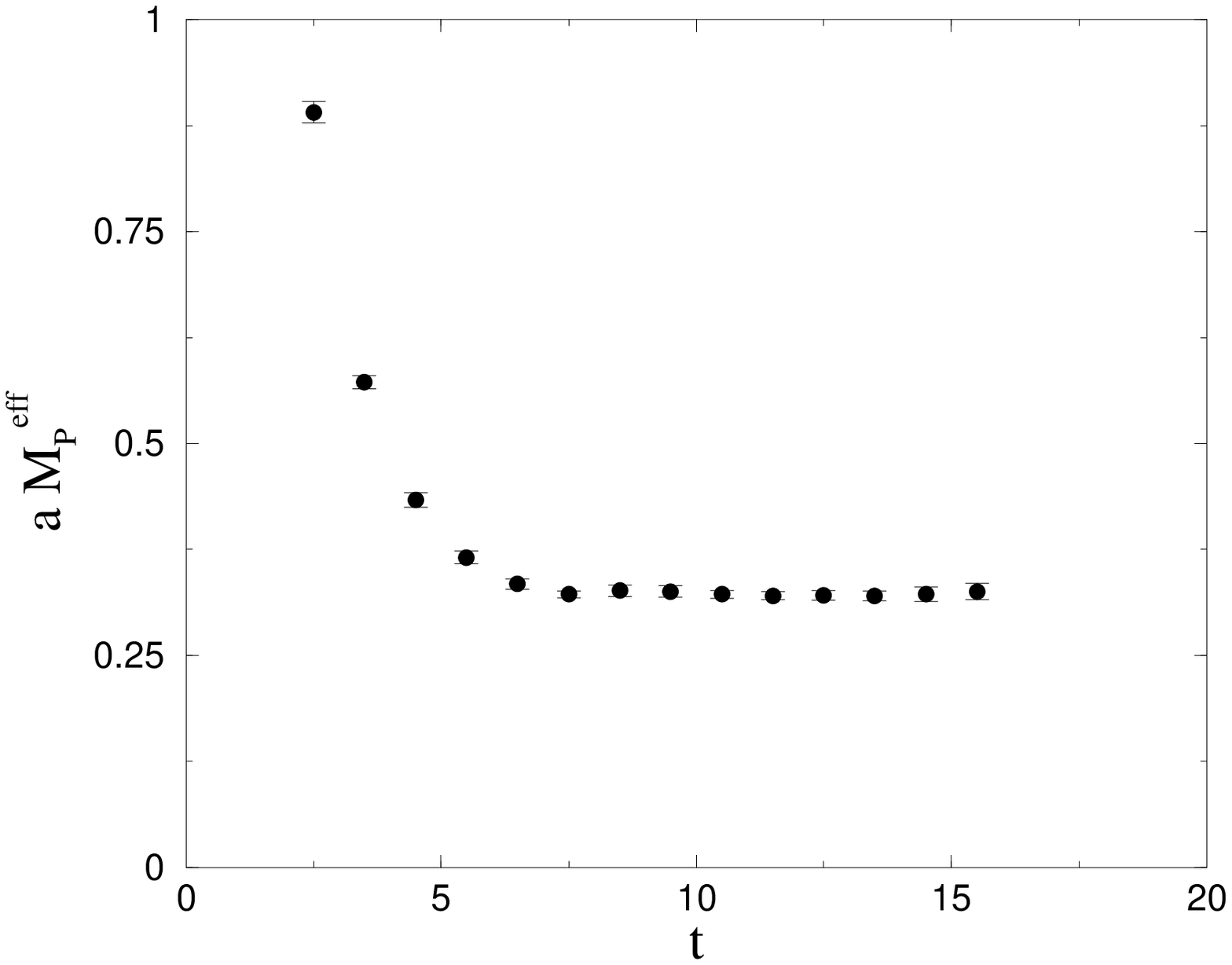,height=6.0cm,width=8.0cm,angle=0}}\\
\end{tabular}
\caption{Left: A representative example ($a m=0.070$) of the effective mass
from $G_{S-P}(t)$. Right: as at left but from $G_{PP}(t)$.
\label{fig:eff_mass}}
\end{center}
\end{figure}
Note that the intercept is compatible with zero. 
This should not come as a surprise
since, even if $A_\mu(x)$ is not the conserved current,
it has the correct behavior under global non-singlet
chiral transformations.

In the quenched approximation the contribution of
exact chiral zero modes of the Neuberger-Dirac operator is not
suppressed by the fermionic determinant.  In some
correlation functions, this can give rise
to large quenched artifacts for small masses.
For example it is easy to show that $G_{PP}(t)$
receives from the unsuppressed zero modes contributions
proportional to $1/m^2$ and $1/m$, which should vanish in the
infinite volume limit, but can be quite sizeable for finite volume.
\begin{table}[htb]
\begin{center}
\begin{tabular}{||c||c|c|c||c|c|c||}
\hline\hline
$a m $ & \multicolumn{3}{c||}{$G_{S-P}$} &
\multicolumn{3}{c||}{$G_{PP}$} \\
\hline
        & $Z_{S-P}$ & $a M_{P}$ & $a f_{P}$ & $Z_{PP}$  &$a M_{P}$ & $a f_{P}$ \\
\hline
0.100   & 0.0040(4) & 0.379(6)  & 0.089(2)  & 0.0042(3) & 0.382(3) & 0.089(2) \\ 
0.085   & 0.0036(4) & 0.348(6)  & 0.085(2)  & 0.0039(3) & 0.352(4) & 0.085(2) \\  
0.070   & 0.0033(4) & 0.315(7)  & 0.081(2)  & 0.0036(3) & 0.321(4) & 0.081(2) \\ 
0.055   & 0.0030(5) & 0.280(9)  & 0.076(2)  & 0.0034(3) & 0.287(5) & 0.077(2) \\ 
0.040   & 0.0026(5) & 0.239(11) & 0.071(2)  & 0.0032(4) & 0.250(7) & 0.073(2) \\ 
\hline
\hline
\end{tabular}
\caption{Mesons masses and matrix elements for all the bare quark masses
considered in the simulations, as obtained from $G_{PP}(t)$
and $G_{S-P}(t)$.\label{tab:masses}}
\end{center}
\end{table}
A clever way to avoid such artifacts has been proposed by the
authors of Ref.~\cite{BNL_spec}, who noticed that the zero modes,
because of their chirality properties, contribute equally
to the $G_{PP}$ and $G_{SS}$ correlation functions, so that
their contributions cancel in the difference
\be\label{eq:BNL_trick}
G_{S-P}(t) = G_{PP}(t) - G_{SS}(t)
\ee
which can also be used to extract the pseudoscalar meson mass
and decay constant,
since the contributions from the heavier scalar mesons
fall off faster.  The drawback is, of course, that the plateau
in the effective mass and, correspondingly, the range that can
be used for the $\cosh$ fit become shorter.  Nevertheless we found
the plateau in the $G_{S-P}$ to be long enough to permit a
meaningful fit.  
\begin{figure}[htb]
\begin{center}
\begin{tabular}{cc}
\mbox{\epsfig{file=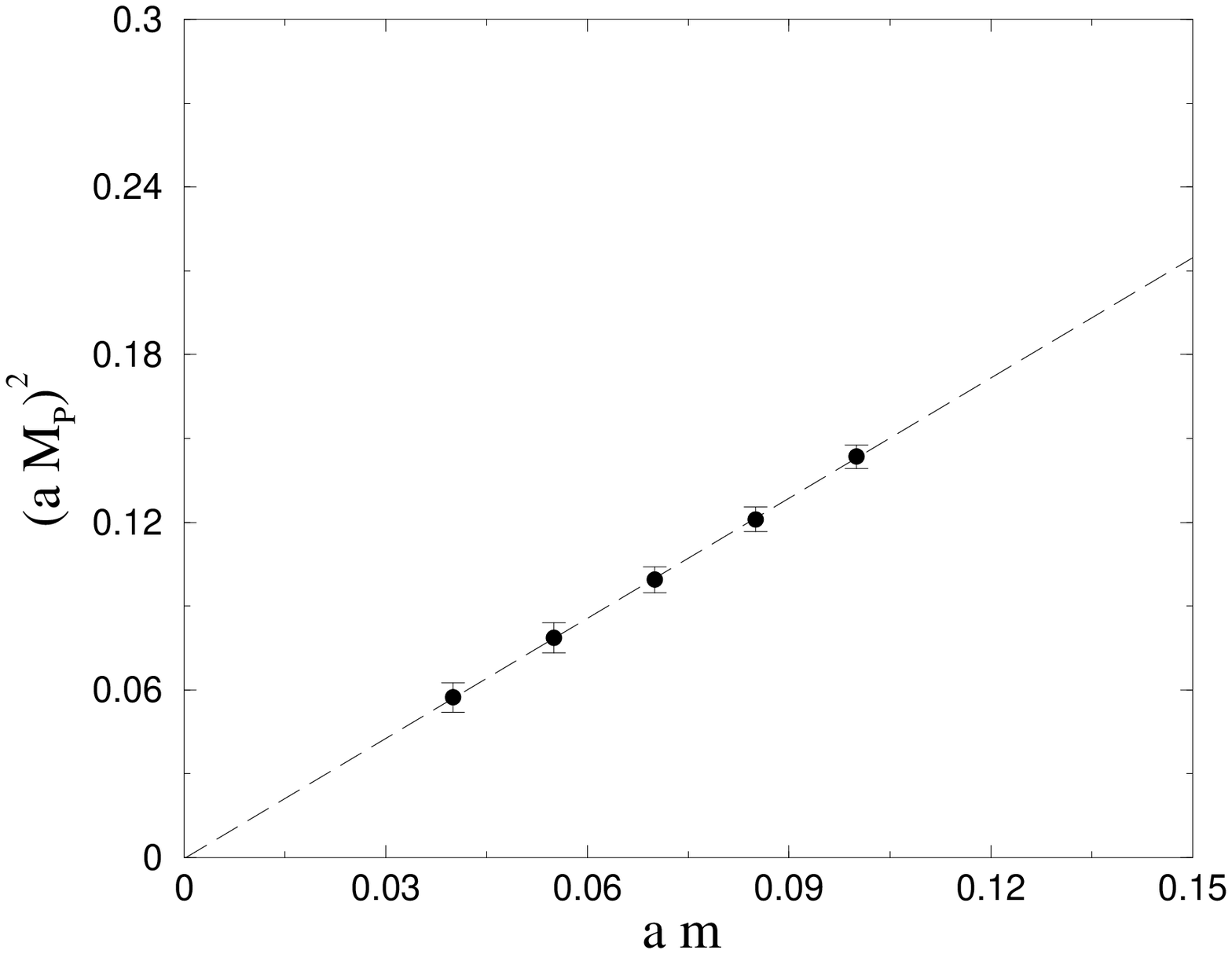,height=6.0cm,width=8.0cm,angle=0}} &
\mbox{\epsfig{file=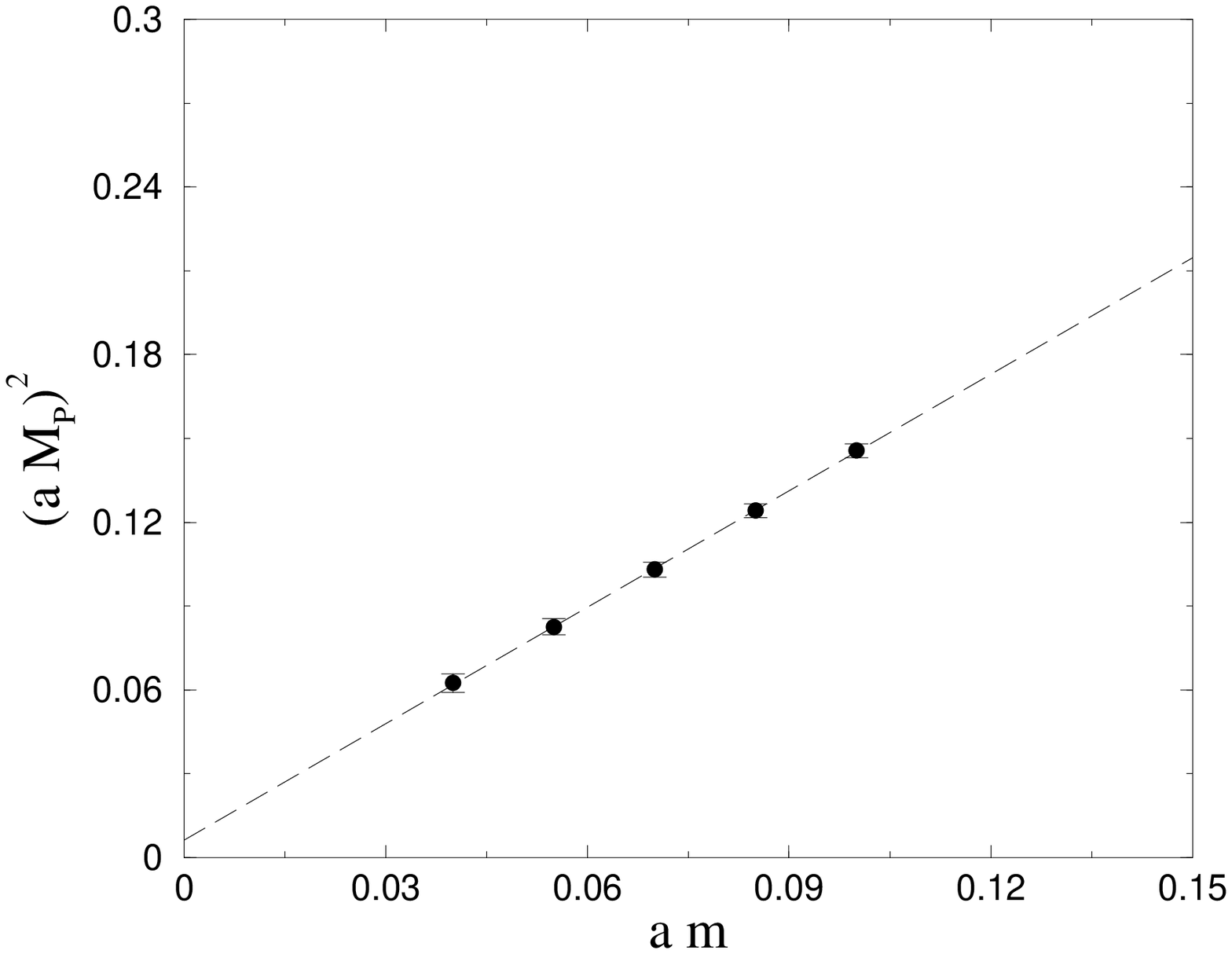,height=6.0cm,width=8.0cm,angle=0}} \\
\end{tabular}
\caption{$(a M_{P})^2$ vs.~$a m$ as obtained from $G_{S-P}(t)$ (left)
and $G_{PP}(t)$ (right). The dashed lines represent the results 
of a linear fit.
\label{fig:mass_extrapo}}
\end{center}
\end{figure}
We fitted $G_{S-P}(t)$
to a single particle propagator with a $\cosh$ dependence on
$t$,
\be
G_{S-P}(t) =  \frac{Z_{S-P}}{a M_{P}}\exp(-\frac{1}{2} a M_{P}T)
\cosh(a M_{P}(\frac{T}{2}-t)) \label{funzfit}
\ee
in the time interval $t_1-t_2=12-16$. The lower limit
was fixed at the point where we found stabilization of
the effective meson masses. The results of the fits are also given
in Table~\ref{tab:masses} and an example of the
effective meson mass from $G_{S-P}(t)$
is shown in first plot of Fig.~\ref{fig:eff_mass}.
We also performed a two $\cosh$ fit of $G_{S-P}(t)$
finding consistent results.

We have also fitted the correlation functions $G_{PP}(t)$ to a single particle
propagator with a $\cosh$ dependence on $t$, as in Eq.~(\ref{funzfit}),
in the time interval $t_1-t_2=10-16$. As before the lower limit
is fixed as the point at which the values of the effective meson
masses become stable. We report our results in Table~\ref{tab:masses}
and in the second plot of Fig.~\ref{fig:eff_mass} we give an example of the
effective meson mass as extracted from $G_{PP}(t)$.

We illustrate in Fig.~\ref{fig:mass_extrapo} the values for $(a M_{P})^2$, 
obtained from $G_{S-P}(t)$ and $G_{PP}(t)$, as a function of the bare quark 
mass $a m$. 
In both cases a linear behavior
\be
(a M_{P})^2 = {\cal A}_{M_P} + {\cal B}_{M_P} (a m)
\ee
fits very well the data with
\be
{\cal A}_{M_P} = -0.0005(68)  \qquad  {\cal B}_{M_P} = 1.43(7)\label{eq:mps_fit2}
\ee
for the masses obtained from $G_{S-P}(t)$ and 
\be
{\cal A}_{M_P} = 0.006(4)    \qquad  {\cal B}_{M_P} = 1.39(3)\label{eq:mps_fit1}
\ee
for those obtained from $G_{PP}(t)$. 
\begin{figure}[htb]
\begin{center}
\begin{tabular}{cc}
\mbox{\epsfig{file=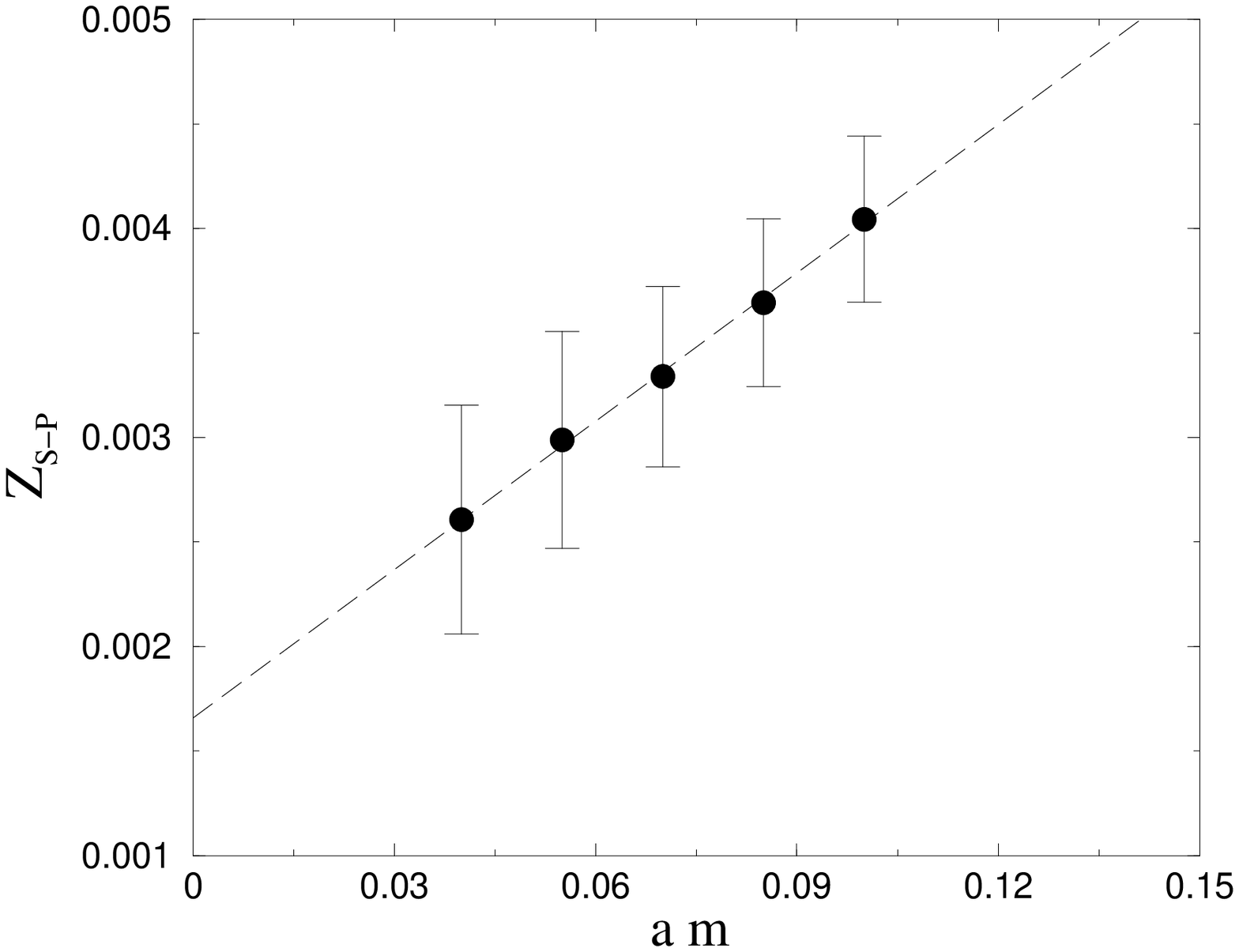,height=6.0cm,width=8.0cm,angle=0}} &
\mbox{\epsfig{file=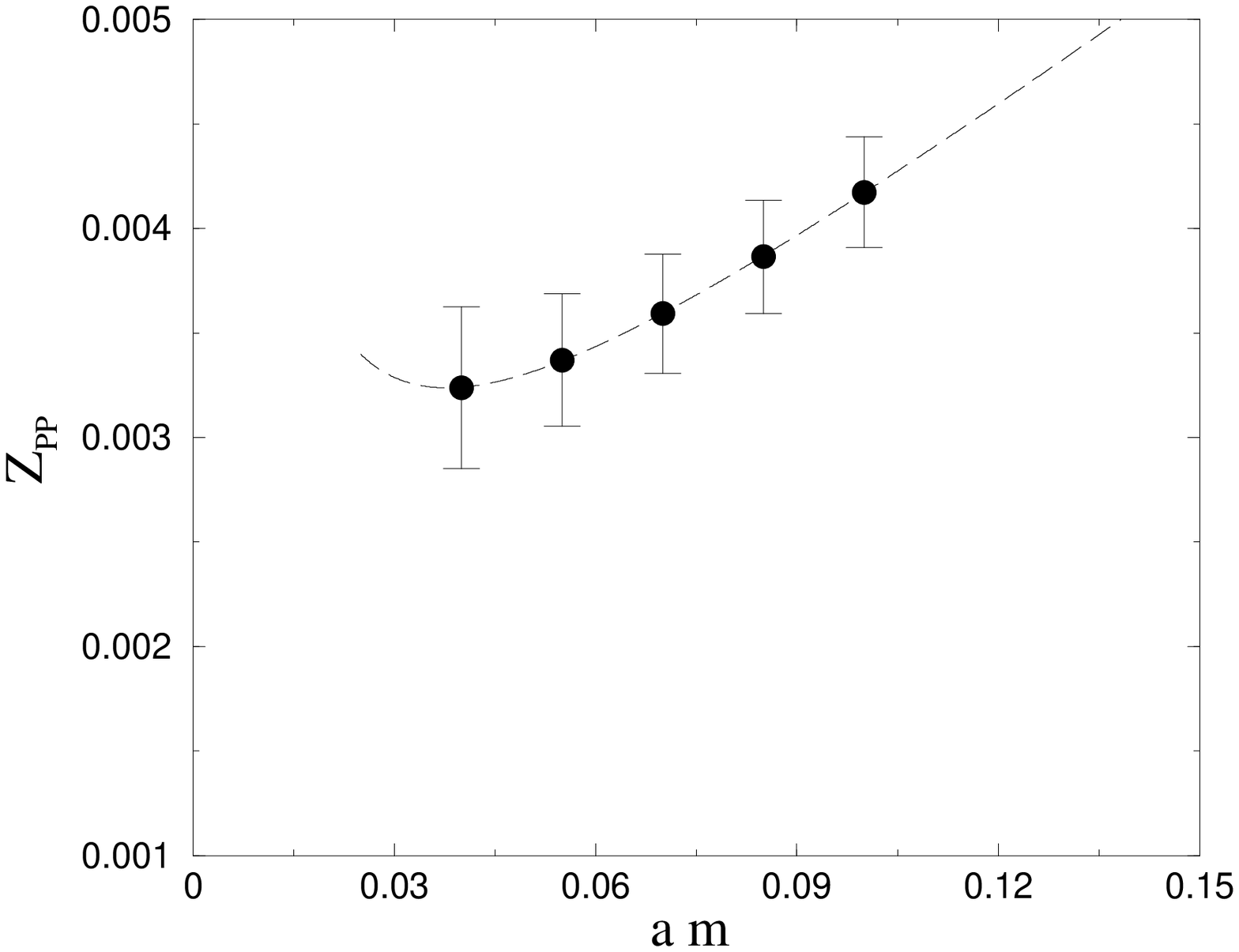,height=6.0cm,width=8.0cm,angle=0}} \\
\end{tabular}
\caption{Left: $Z_{S-P}$ vs.~$(a m)$; the dashed line represents the result
          of a linear fit. Right: $Z_{PP}$ vs.~$(a m)$; the 
          dashed line represents
          a fit of the form of Eq.~(\ref{eq:fitinv}).\label{fig:Z_MQ}}
\end{center}
\end{figure}
We see that the parameters of the two fits are compatible
and that both intercepts vanish within statistical errors.
In particular, one does not notice 
any sign of the singular contributions from zero modes
in the masses obtained
from the pseudoscalar correlation functions.  These are expected to show up
at some point, but one would probably need much higher
statistical accuracy and lower values of $m$ to bring them into evidence. 

Contrary to the pseudoscalar masses, the results for the matrix elements,
parameterized as in Eq.~(\ref{funzfit}) by the factors $Z_{S-P}$ and 
$Z_{PP}$ for $G_{S-P}(t)$ and $G_{PP}(t)$, respectively, 
show more significant differences.
In Fig.~\ref{fig:Z_MQ} we reproduce the results we obtained for $Z_{S-P}$
and $Z_{PP}$ respectively.
We see that, while $Z_{S-P}$ exhibits a linear behavior as function
of $a m$, the graph for $Z_{PP}$ shows a clear indication of
curvature. Linear fits of the form
\be\label{eq:fitq}
Z_i = {\cal A}_i + {\cal B}_i (a m)
\ee
give
\ba
{\cal A}_{S-P} =0.0016(7)  &\qquad& {\cal B}_{S-P} = 0.024(6)\label{eq:fitZ_S-P}\\  
{\cal A}_{PP} = 0.0025(4)  &\qquad& {\cal B}_{PP} =  0.016(3)\label{eq:fitZ_PP}
\ea 
Within the quite large statistical errors, the intercept and the 
slope obtained from $G_{S-P}(t)$ are 
still compatible with those derived from $G_{PP}(t)$.
However the central values are quite different. If we interpret the curvature
in the graph of $Z_{PP}$ as due to pole terms
from the zero modes and try a simple fit of the form 
\be\label{eq:fitinv}
Z_{PP} = {\cal A}_{PP} + {\cal B}_{PP} (a m)+ \frac{{\cal D}_{PP}}{ a m }
\ee
we obtain 
\be\label{eq:fitZ_PPinv}
{\cal A}_{PP} =0.0014(3)  \qquad {\cal B}_{PP} = 0.024(2) 
\qquad {\cal D}_{PP} = 0.000035(16)\; 
\ee 
Now the $\chi^2/\mbox{d.o.f}$ of the fit turns out to be much smaller and the central 
values of intercepts and slope parameters for the fits of $Z_{S-P}$ and $Z_{PP}$ 
are much closer.  The fact that the curvature shows up only
in the results for $Z_{PP}$ points to the fact that what we
are seeing is the effect of the unsuppressed zero modes,
and not of chiral logarithms which would affect both sets
of results (and would most likely become noticeable at much smaller
values of $a m$).  On account of the above, we will
use $G_{S-P}(t)$ to derive our further results.  It must also be said
that some of the observables will be calculated directly at
$m\simeq m_s/2$ (see below), and for these observables
the difference between $Z_{PP}$ and $Z_{S-P}$ is irrelevant within
our statistical error.
\begin{figure}[htb]
\begin{center}
\mbox{\epsfig{file=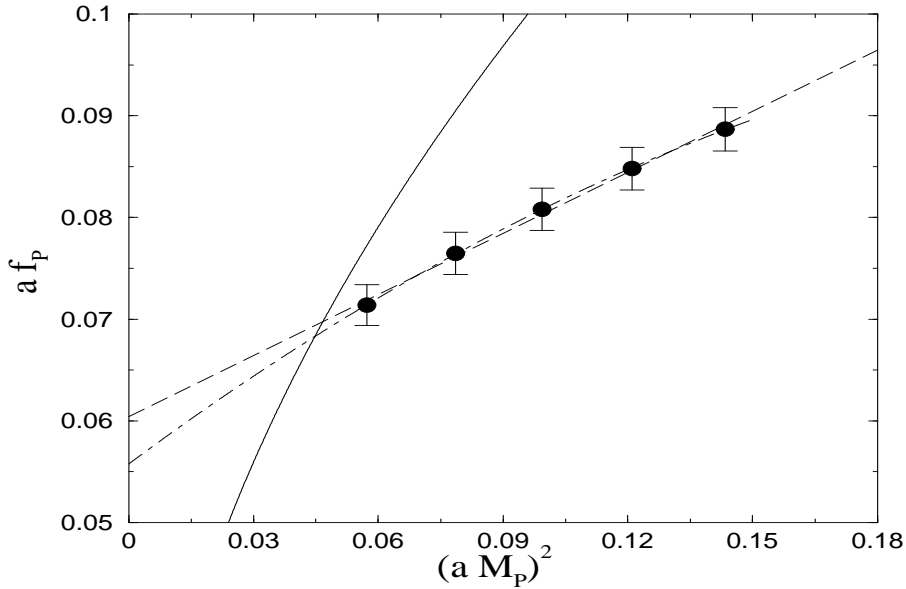,height=8.0cm,width=12.0cm,angle=0}} \\
\caption{The procedure used to derive $af_K$ and $a M_K$.  The dashed
and dashed-dotted lines represent fits to our numerical results.
The solid line represents the curve $(a f_P) = C_{sl} (a M_P)$,
where the coefficient $C_{sl} = f^{exp}_K/M^{exp}_K$. 
\label{fig:fpi_mp^2}}
\end{center}
\end{figure}

The pseudoscalar decay constant is defined through Eq.~(\ref{eq:fp})
and we calculated its value directly from the parameters extracted 
from a $\cosh$ fit to $G_{S-P}(t)$ 
\be
a f_P  =  2 (a m) \frac{\sqrt{Z_{S-P}}}{(a M_P)^2} 
\ee
The values we obtained for $a f_P$ in correspondence
to each simulated mass are reproduced in Table~\ref{tab:masses} 
(where we also reproduce the values derived from $G_{PP}(t)$)
and are shown in Fig.~\ref{fig:fpi_mp^2}. The dashed and the dot-dashed 
lines show the results of a linear and a quadratic fit in $(a M_P)^2$,
respectively. The data slightly favors a quadratic fit, but 
our statistical accuracy is insufficient to rule out
a linear fit as inconsistent with the data.
It is interesting to observe, though, that we obtain
$f_K/f_\pi\simeq 1.14$ and $\simeq 1.23$ from the linear 
and quadratic fits, respectively,
with the results from the quadratic fit in much better
agreement with the experimental value.

Starting from our values for $M_P$ and $f_P$,
we fix the lattice spacing $a^{-1}$~and the physical meson masses
by using the method of ``lattice physical planes'' \cite{lp-method}.
This avoids recourse to a chiral extrapolation for observables
where it is not really needed.
In the plane $[a f_P, (a M_P)^2]$, we plot our lattice data
as well as the curve $(a f_P) = C_{sl} (a M_P)$ 
(the solid line in Fig.~\ref{fig:fpi_mp^2}),
where the coefficient of proportionality $C_{sl}$ ($sl$ for strange-light)
is fixed by 
the experimental value for the
ratio $f^{exp}_K/M^{exp}_K$. \footnote{Throughout the paper we use the following 
experimental numbers: $M^{exp}_K = 495$~MeV, $f^{exp}_K  = 160$~MeV.
We neglect the experimental errors which are well below our statistical 
and systematic errors.}  The point where the two lines meet determines $a f_K$ 
as well as $a M_K$.  Since our statistical accuracy does not allow us to 
discriminate between a linear and quadratic fit to the data for
$a f_P$ vs.~$(a M_P)^2$ , we just use the linear fit
for the tiny extrapolation to the ``Kaon'' point (see  Fig.~\ref{fig:fpi_mp^2}).
Our results are 
\be
a M_K = 0.216(8)  \qquad a f_K = 0.0698(26)
\label{eq:amk}
\ee
(A quadratic fit produces a result which differs from the above by far 
less than the statistical errors). To set the lattice spacing we compare the
value of $a f_K$ with its experimental value and we get
\be
a^{-1}_{f_K} = 2.29(9)
\label{eq:afk}
\ee
This is the value of the lattice spacing we will use throughout
the paper. Again we stress that up to this stage we did not
have to perform any chiral extrapolation, but derived all physical information 
staying close to the region of quark masses used in the actual calculations.
\begin{figure}[htb]
\begin{center}
\mbox{\epsfig{file=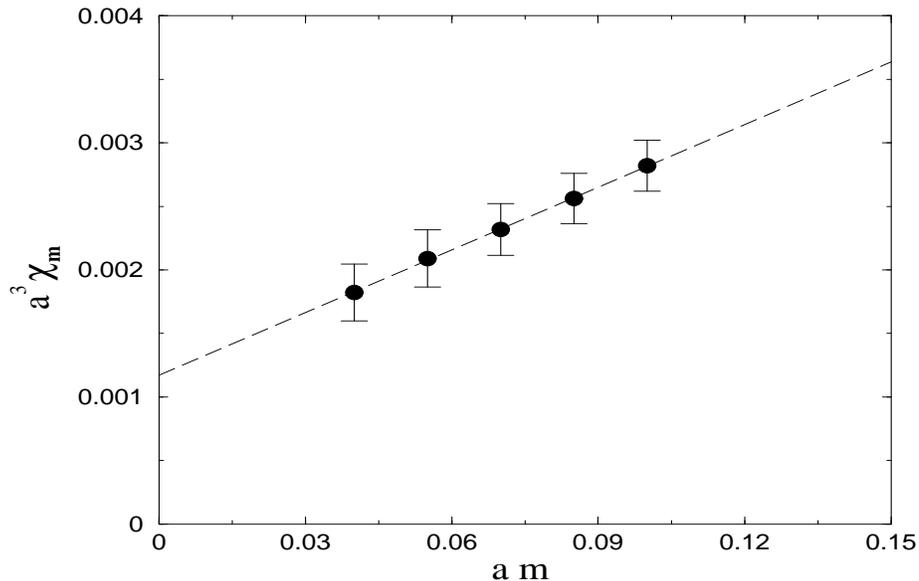,height=8.0cm,width=12.0cm,angle=0}} \\
\caption{The data used to derive the chiral condensate from the GMOR relation.
\label{fig:chiral_cond}}
\end{center}
\end{figure}

From our data, we can also compute the chiral condensate by
using the GMOR relation of
Eq.~(\ref{eq:wip}). In Fig.~\ref{fig:chiral_cond} 
we show the quantity
\be
a^3 \chi_m = -\, (a m) \frac{Z_{S-P}}{(a M_P)^2} 
\ee
as a function of the bare quark mass.
The data exhibits a very good linear behavior and
a linear fit leads to 
\be\label{eq:gmor_numerics}
a^3 \chi = -\, 0.00117(27)
\ee
We performed also a quadratic fit to the data, obtaining
a result consistent with zero for the coefficient of the
quadratic term, but with an error twice as large in the intercept,
because of the additional degree of freedom. Therefore 
we conclude that we see no indication for 
a quadratic term in the fit within our statistical error.  

The chiral condensate can also be computed directly from
Eq.~(\ref{eq:cond_def}). However, away from the chiral limit, 
the r.h.s. of that equation contains power divergent terms (in $a^{-1}$, 
see Eq.~(\ref{eq:cond_div})) and,
in the quenched approximation, infrared divergent 
contributions (in $m$) due to the unsuppressed zero modes.
Therefore quite a severe extrapolation from
our simulated data is needed to determine 
its chiral value. One can take advantage, 
though, of a computational strategy
similar to that used in Eq.~(\ref{eq:BNL_trick}). 
To remove the contribution due to unsuppressed zero modes,
we subtract the scalar-scalar correlation function from the r.h.s of 
Eq.~(\ref{eq:bella}).
 From a quadratic fit in the quark 
mass we obtain
\be
a^3 \chi(a) = -\, 0.00117(42)
\label{eq:chidirect}
\ee
which is in remarkable agreement with Eq.~(\ref{eq:gmor_numerics}).
Of course, the larger error is an indication of the difficulty
of the extrapolation.  Larger statistics and a careful treatment
of the zero mode contributions would be needed
to obtain more reliable and precise results by the direct method. 

\section{Physical Results}\label{sec:Results}
In this section we will use our lattice results to infer the
renormalized values of the sum of the strange and
average up-down quark masses and of the chiral condensate.
To obtain our final values, we still need 
the value of the the scalar renormalization 
constant $Z_S(\mu a)$, which 
we have computed non-perturbatively in the RI/MOM scheme
following the approach proposed in \cite{NP}. The details of these calculations
will be presented in a separate paper \cite{noi_np}.  The result we  
obtained in the $\RI$ scheme is 
\be
Z_S^{\ri}(2\, \mbox{GeV}) = 1.24(5) 
\label{eq:zs}
\ee
where the error is mainly systematics due to the 
chiral extrapolation and to the uncertainty in the value of the 
renormalization scale\footnote{We stress again that 
this number is obtained at $\beta=6.0$ for $\rho=1.4$ and $r=1.0$.}. 
Judging from the comparison of the data at different renormalization 
scales with the logarithmic evolution predicted
by the renormalization group equations at N$^2$LO, 
the discretization errors appear to be well below the error in
Eq.~(\ref{eq:zs}). The result in Eq.~(\ref{eq:zs}) differs by  
about $10\%$ from the bare perturbation theory determination at one loop
$[Z_S^{\ri}(2\, \mbox{GeV})]^{PT}=1.11$ \cite{vicari,capgiu}. However
if we use $\alpha_s^{\msbar}$ as expansion parameter we obtain 
a result consistent within errors with Eq.~(\ref{eq:zs}).
By using N$^2$LO continuum perturbation theory \cite{NNLO_pt}
with $N_f=0$ and $\Lambda_{QCD}=0.238(19)$ \cite{lambdaqcd}
to convert the result in Eq.~(\ref{eq:zs}) into the 
$\MSbar$ scheme, we obtain
\be
Z_S^{\msbar}(2\, \mbox{GeV}) = 1.41(6) 
\label{eq:zsmsbar}
\ee
Had we used the experimental N$^2$LO results
from $\alpha_s(M_Z) = 0.118$, we would have obtained
a value of the scalar renormalization constant $\sim 10\%$
higher than the one given above.  The difference can be taken as an 
indication of the systematic error introduced by the quenched approximation.
Therefore we will add this uncertainty in quadrature to our final estimate of the 
systematic error. N$^3$LO perturbative computations are available 
for the $\RI-\MSbar$ matching of the scalar renormalization 
constant \cite{NNNLO_pt}, but the difference with the N$^2$LO results 
is much below the error induced by the quenching ambiguity discussed above. 
If we had used the procedure proposed recently in Ref.~\cite{HJL_NP2}
on our data, we would have obtained a value for $Z_S^{\msbar}(2\,
\mbox{GeV})$ which agrees within the errors with the one in 
Eq.~(\ref{eq:zsmsbar}), but with a central value $\simeq 10\%$
higher. 

We finally combine this last result with the numbers  
presented in the previous section to obtain our main physical
results. From the value of $a M_K$ in Eq.~(\ref{eq:amk}) and from
the lattice spacing in Eq.~(\ref{eq:afk}), we obtain for the combination
of bare quark masses
\be
m_s(a) + \hat m(a) = 149(9)\, \mbox{MeV}
\label{eq:msbare}
\ee 
where the error is only statistical. Since our volume is fairly large,
we expect our main sources of systematic errors to come from 
discretization effects and from the quenched 
approximation. For a rough estimate of the systematic error due to 
quenching approximation we can use the results in Ref.~\cite{CPPACS_nucleon}.  
From these one sees that, within the quenched approximation, 
the use of different observables to 
calculate the lattice spacing can produce differences in the results 
of $\sim 10\%$.  Had we used $r_0$ \cite{Guagnelli:1998ud} 
to fix our lattice spacing   
we would have obtained a number $\sim 7\%$ higher than the one in
Eq.~(\ref{eq:msbare}).  Combining this fact with the
results in \cite{Garden:2000fg}, we would infer 
that discretization uncertainties are below $10 \%$.  
In order to be conservative, we will take $15 \%$ as the estimate
of our overall systematic error in the renormalized quark masses 
due to quenching and discretization effects. A more precise estimate 
of the systematic errors will need much more extensive simulations, 
which at present would
be beyond our capability and the exploratory scope of this 
work.
Combining the results in Eqs.~(\ref{eq:zsmsbar}) and~(\ref{eq:msbare}) we obtain
\be
(m_s + \hat m)^{\ri}(\mbox{2 GeV}) = 120 \pm 7 \pm 21\, \mbox{MeV}
\label{eq:ml+ms}
\ee
which represents one of the major results of this paper. By using
Eqs.~(\ref{eq:chipt}) and (\ref{eq:zsmsbar}) 
the above translates to 
\be
m_s^{\msbar}(\mbox{2 GeV}) = 102 \pm 6 \pm 18\; \, \mbox{MeV} 
\ee
This result agrees very well with the current lattice world 
average \cite{vittorio_rev}. 

Insofar as the value of the condensate is concerned,
if we used the standard two-step approach, i.e.~first measure the 
dimensionless condensate, see Eqs.~(\ref{eq:gmor_numerics}) 
or~(\ref{eq:chidirect}), and then
multiply it by the cubic power of the lattice spacing,  
the result would be affected by a very large 
systematic error due to the uncertainty in the determination of the 
lattice spacing in quenched simulations. 
Instead, we will use an alternative method \cite{Gimenez:1999uv}. We write the GMOR 
relation (\ref{eq:GMOR_true}) for the
renormalized condensate as follows
\be
\chi(a)
= - \dfrac{1}{4} f_\chi^2 {\cal B}_{M_P} a^{-1}
\label{eq:psi1}
\ee
where ${\cal B}_{M_P}$ is defined in Eq.~(\ref{eq:mps_fit2}) 
and
$f_\chi=0.1282$~GeV is the ``experimental'' value, in physical units,
of the pseudoscalar decay constant extrapolated to the chiral limit.
While computing the condensate from the
above formula relies on an additional element of experimental
information, it has several advantages. The most important is that, by
expressing the condensate in terms of $f_\chi$, 
we are left with only one power of
the UV cutoff $a^{-1}$.
Another advantage is that, if we assume that the relation between
$M_P$ and $m$ stays substantially linear for a range of quark
masses extending to $\sim m_s$ (and our numerical results
validate this, see Fig.~\ref{fig:mass_extrapo}), then 
there is no need for an extrapolation to the chiral limit
to evaluate ${\cal B}_{M_P}$.
With this method we obtain
\be\label{eq:cond_best}
\langle\bar \psi \psi\rangle^{\ri}(\mbox{2 GeV}) = -\, 0.0167 \pm 0.0010 \pm
0.0029 \, \mbox{GeV}^3
\ee
where the estimate of the systematic error has been made using the
same criteria we used for estimate of the error in the quark mass.
This is our best value for the chiral condensate.
It is interesting to note that, if we had used the standard technique,
starting from Eq.~(\ref{eq:gmor_numerics}) and with the lattice spacing
of Eq.~(\ref{eq:afk}), we would have obtained 
a result with a central value very close to the value in Eq.~(\ref{eq:cond_best}),
but with a much higher systematic error.

Finally, from Eq.~(\ref{eq:cond_best}) and N$^2$LO matching, 
we get
\be\label{eq:cond_msbar}
\langle\bar \psi \psi\rangle^{\msbar}(\mbox{2 GeV}) = -\, 0.0190 \pm 0.0011 \pm 0.0033\, \mbox{GeV}^3 = 
-\, \left[267 \pm 5 \pm 15 \, \mbox{MeV}\right]^3
\ee
This result is in very good agreement with the result obtained by the authors 
of Refs.~\cite{HJL_NP1,HJL_NP2}, while it is smaller than the result 
in Ref.~\cite{DeGrand:2001ie}, even if still compatible within 
errors. Our result is also compatible
within errors with the number obtained few years ago 
in Ref.~\cite{Gimenez:1999uv} with Wilson-type fermions. 
We expect, though,
the systematics due to the discretization effects to be smaller 
(${\cal O}(a^2)$) in the result reported in 
Eq.~(\ref{eq:cond_msbar}) than 
the error (${\cal O}(a)$) which affects the determination 
in Ref.~\cite{Gimenez:1999uv}.  

\section{Conclusions}\label{sec:conclusions}
In the overlap regularization chiral symmetry is preserved
at finite lattice spacing and finite volume, 
therefore there is no mixing among operators of different chirality.
As a consequence no additive quark mass renormalization is required and
no fine tuned parameters are needed to compute ${\cal O}(a)$
improved masses and matrix elements.  Our results have indeed 
produced a remarkable verification of ``good chiral behavior''
both in the axial Ward identity and in the pseudoscalar masses.

In this paper we presented the results of the first computation
of $(m_s+\hat m)$ with overlap fermions in the quenched approximation.
To avoid uncertainties due to lattice perturbation
theory, we computed the multiplicative renormalization constant 
$Z_S(\mu a)$ non-perturbatively in the RI/MOM scheme.
Our main results have been summarized in the
introduction.  While the systematics errors due to quenching are common
to previous calculations, the other systematic errors
(mostly discretization effects) are different than with other lattice
regularizations and likely to be smaller, because of chiral symmetry.
We also computed the chiral condensate $\langle \bar \psi \psi \rangle$
from the GMOR relation and directly. Even if, given our statistical 
and systematic errors, the former method is more reliable, it is 
rewarding to notice that the two determinations are in good agreement.

The calculation of light quark masses uses many of the ingredients 
needed for a lattice calculation of weak matrix elements, although
the latter is computationally more demanding.  From this point
of view, the very good agreement between our results for the quark
masses and the current lattice world average
bodes well for the use of the overlap formalism also for matrix
element calculations.  Our investigation has been
mostly of exploratory nature.  One would need to extend it to
larger volumes and better statistics.  One should also find
a more direct way to isolate and account for the effects of the zero
modes.  Nevertheless, we believe that it gave a strong indication that 
the overlap formalism can be used effectively, with known algorithms and the
present generation of computers, for large scale QCD calculations, at
least in the quenched approximation.  Thus we would conclude that
it represents a very promising non-perturbative
regularization for solving long standing problems, such as the proof
of the $\Delta I =1/2$ rule and the calculation of
$\epsilon'/\epsilon$, which would be hard to address with conventional
regularizations.

\section*{Acknowledgments}
We wish to thank Boston University's Center for Computational Science and 
Office of Information Technology at Boston University for generous
allocations of supercomputer time and the Scientific Computing
and Visualization group for invaluable technical assistance.
We also gratefully acknowledge the use of the gauge configurations produced by
the authors in Ref.~\cite{GaugeConn}.  This work has been supported in part 
under DOE grant DE-FG02-91ER40676.

\end{document}